\documentclass[equations,12pt]{article}
\usepackage{amsfonts}
\usepackage{a4,tikz}
\usepackage{geometry}
\usepackage{color} 
\usepackage{subfigure}
\makeatletter
\newcommand\ackname{Acknowledgements}
\if@titlepage
  \newenvironment{acknowledgements}{%
      \titlepage
      \null\vfil
      \@beginparpenalty\@lowpenalty
      \begin{center}%
        \bfseries \ackname
        \@endparpenalty\@M1
      \end{center}}%
     {\par\vfil\null\endtitlepage}
\else
  \newenvironment{acknowledgements}{%
      \if@twocolumn
        \section*{\abstractname}%
      \else
        \small
        \begin{center}%
          {\bfseries \ackname\vspace{-.5em}\vspace{\z@}}%
        \end{center}%
        \quotation
      \fi}
      {\if@twocolumn\else\endquotation\fi}
\fi
\makeatother
\usepackage{centernot}
\usepackage{mathtools}
\usepackage{ stmaryrd }
\usepackage{amsmath}
\usepackage{amssymb}
\usepackage{amsthm}
\usepackage{mathrsfs}
\usepackage{eucal}
 \usepackage[usenames,dvipsnames]{pstricks}
 \usepackage{epsfig}
 \usepackage{pst-grad} 
 \usepackage{pst-plot} 
\renewcommand{\theequation}{\arabic{equation}}
\usepackage{amssymb,amsmath}
\usepackage{color}
\usepackage{accents}
\usepackage{texdraw}
\usepackage{eucal}
\usepackage{epic,epsfig}
\usepackage{graphicx}

\theoremstyle{definition}

\numberwithin{equation}{section}
\DeclareMathAccent{\wtilde}{\mathord}{largesymbols}{"65}
\DeclareMathAccent{\what}{\mathord}{largesymbols}{"62}

\def\m@th{\mathsurround=0pt}
\mathchardef\bracell="0365
\def\upbrall{$\m@th\bracell$}
\def\undertilde#1{\mathop{\vtop{\ialign{##\crcr
    $\hfil\displaystyle{#1}\hfil$\crcr
     \noalign
     {\kern1.5pt\nointerlineskip}
     \upbrall\crcr\noalign{\kern1pt
   }}}}\limits}
\def\m@th{\mathsurround=0pt}
\mathchardef\bracell="0365
\def\upbrall{$\m@th\bracell$}
\def\underhat#1{\mathop{\vtop{\ialign{##\crcr
    $\hfil\displaystyle{#1}\hfil$\crcr
     \noalign
     {\kern1.5pt\nointerlineskip}
     \upbrall\crcr\noalign{\kern1pt
   }}}}\limits}
\usepackage{pgf}
\usepackage{tikz}
\usepackage{verbatim}
\usepackage[utf8]{inputenc}
\usetikzlibrary{arrows,automata}
\usetikzlibrary{positioning}

\def\theequation{\arabic{section}.\arabic{equation}}

\newcommand{\bblu}{\begin{color}{blue}}
\newcommand{\bred}{\begin{color}{red}}
\newcommand{\ecl}{\end{color}}

\newcommand{\be}{\begin{equation}}
\newcommand{\ee}{\end{equation}}
\newcommand{\bea}{\begin{eqnarray}}
\newcommand{\eea}{\end{eqnarray}}
\newcommand{\bse}{\begin{subequations}}
\newcommand{\ese}{\end{subequations}}
\newcommand{\nn}{\nonumber}

\begin{document}

\def\theequation{\arabic{section}.\arabic{equation}}

\newtheorem{thm}{Theorem}[section]
\newtheorem{lem}{Lemma}[section]
\newtheorem{defn}{Definition}[section]
\newtheorem{ex}{Example}[section]
\newtheorem{rem}{}
\newtheorem{criteria}{Criteria}[section]
\newcommand{\ra}{\rangle}
\newcommand{\la}{\langle}
\everymath{\displaystyle}

\title{\center{\textbf{\Large{On the multiplicative form of the Lagrangian}}}}
\author{\\\\Kittikun Surawuttinack$^a$, Sikarin Yoo-Kong$^{a,b,c } \thanks{Corresponding author, email: syookong@gmail.com}$ and Monsit Tanasittikosol$^{a} $ \\
\small $^a $\emph{Theoretical and Computational Physics (TCP) Group, Department of Physics,}\\ 
\small \emph{Faculty of Science, King Mongkut's University of Technology Thonburi, Bangkok 10140, Thailand.}\\
\small $^b $\emph{Theoretical and Computational Science Centre, Faculty of Science,}\\ 
\small \emph{ King Mongkut's University of Technology Thonburi, Bangkok 10140, Thailand.}\\
\small $^c $ \emph{Ratchaburi Campus, King Mongkut's University of Technology Thonburi, Ratchaburi, 70510, Thailand}\\
}
\maketitle

\abstract
An alternative class of the Lagrangian called the multiplicative form is successfully derived for a system with one degree of freedom for both non-relativistic and relativistic cases. This new Lagrangian can be considered as a 1-parameter: $\lambda$ extended class from the standard additive form of the Lagrangian since both yield the same equation of motion. Remarkably, the multiplicative form of the Lagrangian could be treated as a generating function to produce an infinite hierarchy of Lagrangians which give the same equation of motion. This nontrivial set of Lagrangians confirms that indeed Lagrange function is not unique.

\section{Introduction}\label{intro}
\setcounter{equation}{0}
The Lagrangian formalism is one of the standard methods in classical mechanics. The Lagrange function and its associated equations of motion are expressed in terms of generalised coordinates. The advantage of this method is that it helps identifying the conserved quantities through the cyclic variables. Furthermore, one may employ the Noether's theorem to find conservation laws. The form of the Lagrangian can be expressed in terms of energy scalar functions which, in this paper, is called a standard additive form. Using Legendre transformation, one can construct the Hamilton function and we can study the system from the Hamiltonian formalism point of view.

It is commonly well known that the Lagrangian possesses a non-uniqueness property.  We can multiply a constant $\beta$ or add up a constant $\alpha$ to the Lagrangian: $L\rightarrow \beta L+\alpha$, leaving no effect to the equation of motion of the system because the Euler-Lagrange equations involve only derivatives of the Lagrangian. We also can add the total derivative term to the Lagrangian: $L\rightarrow L+\frac{df(x,t)}{dt}$, affecting nothing to the equations of motion of the system.

An interesting point is the following. With the given equations of motion, we must first find the corresponding Lagrangian if we would like to work in the Lagrangian formalism. Some attempts had been done in \cite{Ari} (and references therein) to solve for the Lagrangian, but the general mechanism is still open. In this paper, we ask the following questions: Is there an alternative form of the Lagrangian, as well as its associated Hamiltonian, apart from the standard additive one? And if so, can they be solved systematically?

The organisation of the paper is as follows. In section \ref{ML}, we set out to find the multiplicative Lagrangian for the non-relativistic case with one degree of freedom. Then we establish the Legendre transformation between multiplicative Hamiltonian and multiplicative Lagrangian. We also solve the multiplicative Hamiltonian and compare the result with the Legendre transformation. An interesting observation of the multiplicative form of the Lagrangian will be made as a generating function for an infinite hierarchy of Lagrangians.
In section \ref{RL}, we extend the idea to the relativistic case to solve for the multiplicative Lagrangian and multiplicative Hamiltonian. The infinite hierarchy of Lagrangians is also considered.
 Finally in the last section, we give the summary with some remarks.
\section{Non-relativistic case}\label{ML}
In this section, we are interested a system with one degree of freedom and the equation of motion is given by
\begin{equation}\label{NE}
m\ddot{x}=-\frac{dV(x)}{dx}\;,
\end{equation}
where $V(x)$ is the potential. The Lagrangian can be written in the form
\begin{equation}\label{LN}
L_N(x,\dot x)=T(\dot x)-V(x)\;,
\end{equation}
where $T(\dot x)=m\dot x^2/2$ which is the kinetic energy. Using Legendre transformation, we obtain the Hamiltonian
\begin{equation}\label{HN}
H_N(p,x)=p\dot x-L(x,\dot x)=T(p)+V(x)\;,
\end{equation}
where $T(p)=p^2/2m$ which is the kinetic energy, but in terms of momentum variable: $p=m\dot x$. Equation \eqref{NE} then can be written in the form
\begin{equation}\label{PE}
\dot{p}=-\frac{dV(x)}{dx}\;.
\end{equation}
We now ask whether there exists an alternative form of the Lagrangian as well as the Hamiltonian to produce the same equation of motion. 
\subsection{Multiplicative Lagrangian}
To answer the question, we start with an Ansatz form\footnote{This form is inspired by the result in \cite{DR}.} of the Lagrangian $L=F(\dot{x}) G(x)$, where $F$ and $G$ are yet to be determined. We now find that the action functional is
\begin{eqnarray} 
S[x] &=& \int_0^T  L (\dot x,x)dt
= \int_0^T F(\dot{x})G(x) dt~.
\end{eqnarray}
According to the variational principle, we find that
\begin{eqnarray} \label{EL1}
S [x+\delta x] &=& \int_0^T L(\dot x+\delta\dot x,x+\delta x)dt \nn
= \int_0^T F(\dot x+\delta\dot x)\ G(x+\delta x) dt \nn \\
S+ \delta S+... &=&  \int_0^T  \left(\left[ F(\dot{x})+\delta\dot x\frac{dF(\dot{x})}{d\dot{x}}+...\right] \left[G(x)+\delta x\frac{dG(x)}{dx}+...\right]\right) dt \nn\\
&=& \int_0^T \left(F(\dot{x})G(x)+  \delta\dot x\frac{dF(\dot{x})}{d\dot{x}} G(x) + \delta xF(\dot{x})\frac{dG(x)}{dx}+... \right) dt \nn\\
\delta S&=& \int_0^T \left( \delta\dot x\frac{dF(\dot{x})}{d\dot{x}} G(x) + \delta xF(\dot{x})\frac{dG(x)}{dx}  \right) dt \nn\\
&=& \int_0^T\left(  - \frac{d}{dt}\left( G(x) \frac{d F(\dot{x})}{d \dot{x}} \right) + F(\dot{x})  \frac{d G(x)}{d x}  \right)\delta x dt\;,
\end{eqnarray}
with the use of $\delta x(0)=\delta x(T)=0$. According to \eqref{EL1}, $\delta S$ vanishes if 
\begin{eqnarray}\label{EL2}
 F(\dot{x})  \frac{dG(x)}{d x}   - \frac{d}{dt}\left( G(x) \frac{dF(\dot{x})}{d \dot{x}} \right)  = 0\;,
\end{eqnarray}
which could be treated as a new Euler-Lagrange equation associated with the multiplicative Lagrangian. Furthermore, \eqref{EL2} can be re-written in the form
\begin{eqnarray}\label{AA1}
\frac{1}{\ddot{x}G} \frac{d G}{d x}\left( F - \dot{x}\frac{dF}{d \dot{x}}\right) - \frac{d^2 F}{d\dot{x}^2} &=& 0 \;.
\end{eqnarray}
Using the technique of separation of variables, we set
\begin{equation} \label{EG}
 \frac{1}{\ddot{x}G} \frac{dG}{dx}  = A \Rightarrow
\frac{1}{G}\frac{d G}{d x} = A\ddot{x}\;.
 \end{equation}
Using equation of motion \eqref{NE},
 \eqref{EG} becomes
\begin{equation} \label{EG2}
\frac{1}{G}\frac{d G}{d x} 
= -\frac{A}{m}\frac{d V(x)}{d x }~,
 \end{equation}
The parameter $A$ is a constant which will be determined later.  The solution of  \eqref{EG} is simply taken the form of
\begin{equation}\label{G1}
G(x) = \alpha_1 e^{-\frac{AV(x)}{m}}\;,
\end{equation}
where $\alpha_1$  is a constant. Using  \eqref{EG},  \eqref{AA1} now reduces to
\begin{equation} 
A \left( F - \dot{x}\frac{d F}{d x}\right) - \frac{d^2 F}{d \dot{x}^2} = 0\;,
\end{equation}
 and the solution for $F$ takes the form
\begin{eqnarray} \label{F1}
F(\dot{x}) = \alpha_2\dot{x} - \alpha_3 \left( e^{-\frac{A\dot{x}^2}{2}} + \dot{x}A \int_0^{\dot x} e^{-\frac{Av^2}{2}}dv\right )\;,
\end{eqnarray} 
where $\alpha_2$ and $\alpha_3$ are constant. \\\\
\emph{Remark: Equation \eqref{AA1} can be modified as follows
\begin{equation}
\frac{1}{G} \frac{d G}{d x}\left( F - \dot{x}\frac{dF}{d \dot{x}}\right) - \frac{1}{m}\frac{d^2 F}{d\dot{x}^2} \frac{dV}{dx}=- \frac{d^2 F}{d\dot{x}^2}\left(\ddot x+\frac{1}{m} \frac{dV}{dx}\right)\equiv A(x,\dot x)\;. \nn
\end{equation} 
The condition $A(x,\dot x)=0$ for arbitrary independent variables $x$ and $\dot x$ must hold for equations of motion for different Lagrangians to coincide, resulting again to \eqref{G1} and \eqref{F1}.
}
\\
\\
Using the results of $F$ and $G$, we then obtain the multiplicative form of the Lagrangian which is given by
\begin{equation} 
L(x,\dot x)  =   \left[ k_1\dot{x} - k_2 \left( e^{-\frac{A\dot{x}^2}{2}} + \dot{x}A \int_0^{\dot x} e^{-\frac{Av^2}{2}}dv \right) \right]e^{-\frac{AV(x)}{m}}\;, \label{AA2}
\end{equation}
where $k_1=\alpha_1\alpha_2$ and $k_2=\alpha_1\alpha_3$ are constant and yet to be determined. In order to identify all the remaining constants, we expand the exponential terms
\begin{eqnarray}\label{LTT}
L &=& \left( k_1\dot x - k_2\left[1-\frac{A\dot{x}^2}{2} + \frac{1}{2!}{\left(\frac{A\dot{x}^2}{2}\right)}^2 - \ldots\right] - k_2 \dot{x}A \int_0^{\dot x} \left[1-\frac{Av^2}{2} + \frac{1}{2!}{\left(\frac{Av^2}{2}\right)}^2 \right.\right.\nn\\
&&\left.\left.- \ldots\phantom{\frac{1}{2}}\right]dv \right)  \left[ 1 - \frac{AV(x)}{m} +\frac{1}{2!}{\left(\frac{AV(x)}{m}\right)}^2 - \ldots \right]\;.
\end{eqnarray}
It is found that if we take $A$ to be an inverse square of the velocity: $\lambda^{-2}$, $k_1$ to be zero and  $k_2$ to be in energy unit: $-m\lambda^2$, the Lagrangian \eqref{LTT} in the limit that $\lambda$ approaches to infinity
\begin{eqnarray} \label{AA3}
 \lim_{\lambda \rightarrow \infty }L &=& \lim_{\lambda  \rightarrow \infty} \left( m\lambda^2\left[1 - \frac{\dot{x}^2}{2\lambda^2} + \ldots \right] +m\dot{x} \int_0^{\dot x}\left[1-\frac{v^2}{2\lambda^2} + \ldots\right]d v \right) \nn\\
&&\times \left( 1 -\frac{V(x)}{m\lambda^2} + \ldots \right)\nn \\ 
&=& \lim_{\lambda \rightarrow \infty} m\lambda^2 \left(  1 + \frac{{\dot{x}}^2}{2\lambda^2} + \ldots \right)\left( 1 - \frac{V(x)}{m\lambda^2} + \ldots \right)\nn\\
 &=& \lim_{\lambda \rightarrow \infty} m\lambda^2 \left( 1 +\frac{{\dot{x}}^2}{2\lambda^2} - \frac{V(x)}{m\lambda^2}  +    \ldots \right) \nn\\
\lim_{\lambda \rightarrow \infty }(L-m\lambda^2) &=&  \frac{m\dot x^2}{2}-V(x)=L_N\;,
\end{eqnarray}\\
where $L_N$ is the standard Lagrangian in the additive form \eqref{LN}. Thus the Lagrangian takes the form of
\begin{equation}
L_\lambda(x,\dot x) = m\lambda^2\left( e^{-\frac{\dot{x}^2}{2\lambda^2}} + \frac{\dot{x}}{\lambda^2} \int_0^{\dot x} e^{-\frac{v^2}{2\lambda^2}} d v \right)e^{-\frac{V(x)}{m\lambda^2}}\;, \;\; \;  \label{SS1}
\end{equation}
which can be treated as the 1-parameter, namely $\lambda$, extended class of the Lagrangian. Indeed, Lagrangian (2.13) leads to the standard Lagrangian in the limit case (2.12).
\\\\
Next, the momentum is given by
\begin{equation}\label{pc}
p_\lambda= \frac{\partial L_\lambda}{\partial \dot{x}}= m  \left( \int_0^{\dot x} e^{-\frac{v^2}{2\lambda^2}} d v \right)e^{-\frac{V(x)}{m\lambda^2}} \; ,
\end{equation}
which in the limit that $\lambda$ approaches to infinity. we recover the standard momentum: $
\lim_{\lambda \rightarrow \infty }p_\lambda=p=m\dot x$. 
\\\\
Next, we find that
\begin{eqnarray}
\frac{\partial L_\lambda}{\partial x} &=&   \left( e^{-\frac{\dot{x}^2}{2\lambda^2}} + \frac{\dot{x}}{\lambda^2} \int_0^{\dot x} e^{-\frac{v^2}{2\lambda^2}} d v  \right)e^{-\frac{V(x)}{m\lambda^2}} \left( -\frac{d V(x)}{d x} \right) \; ,\label{a1}  \\
\frac{\partial^2 L_\lambda}{\partial x\partial\dot{x}} &=& \frac{1}{\lambda^2}  \left( \int_0^{\dot x} e^{-\frac{v^2}{2\lambda^2}} d v  \right) e^{-\frac{V(x)}{m\lambda^2}} \left(- \frac{d V(x)}{d x} \right)\; .\label{a2}
\end{eqnarray}
Substituting \eqref{a1} and \eqref{a2} into the Euler-Lagrange equation,
$$ \frac{\partial L_\lambda}{\partial x} -\frac{d}{dt}\left( \frac{\partial L_\lambda}{\partial \dot{x}} \right) = 0 \; , $$  
with the help of the relation
\begin{equation}
\frac{d}{dt} \left( \frac{\partial L_\lambda}{\partial \dot{x}}\right)= \ddot{x}\frac{\partial^2 L_\lambda}{\partial \dot{x}^2} + \dot{x}\frac{\partial^2 L_\lambda}{\partial x \partial \dot {x}}~,
\end{equation}
we obtain 
\begin{eqnarray}\label{DCS}
0 &=& \left( e^{-\frac{\dot{x}^2}{2\lambda^2}} + \frac{\dot{x}}{\lambda^2} \int_0^{\dot x} e^{-\frac{v^2}{2\lambda^2}} d v  \right)e^{-\frac{V(x)}{m\lambda^2}} \left( -\frac{d V(x)}{d x} \right)  -m\ddot{x}  \left( e^{-\frac{\dot{x}^2}{2\lambda^2}}  \right)e^{-\frac{V(x)}{m\lambda^2}} \nn \\
&&- \frac{\dot{x}}{\lambda^2}  \left( \int_0^{\dot x} e^{-\frac{v^2}{2\lambda^2}} d v  \right) e^{-\frac{V(x)}{m\lambda^2}} \left(- \frac{ d V(x)}{d x} \right) \nn  \\
&=& e^{-\frac{\dot{x}^2}{2\lambda^2}} \left( - \frac{d V(x)}{d x} - m\ddot{x}\right). 
\end{eqnarray}
Equation \eqref{DCS} implies the equation of motion \eqref{NE}
which completes the quest for searching an alternative class of the Lagrangian.

\subsection{Multiplicative Hamiltonian}\label{MH}
In the previous subsection, the multiplicative form of the Lagrangian was established. In the additive case, the Hamiltonian and the Lagrangian are connected through Legendre transformation. 
Here it is interesting to see how it works in the case of the multiplicative Lagrangian. According to the Neother's theorem, we also find that
\begin{eqnarray}
\frac{dL_\lambda}{dt}&=&\frac{\partial L_\lambda}{\partial x}\dot x+\frac{\partial L_\lambda}{\partial \dot x}\ddot x\nn\\
&=&\frac{d}{dt}\left(\frac{\partial L_\lambda}{\partial \dot x}\right)\dot x+\frac{\partial L_\lambda}{\partial \dot x}\ddot x\nn\\
\Rightarrow 0&=&\frac{d}{dt}\left(\frac{\partial L_\lambda}{\partial \dot{x}} \dot{x} - L_\lambda(\dot x,x) \right)\;,
\end{eqnarray}
which implies that the terms in the bracket must be invariant in time leading to
\begin{equation}\label{HC2}
H_\lambda(p,x) \equiv \frac{\partial L_\lambda}{\partial \dot{x}} \dot{x} - L_\lambda(\dot x,x)\;,
\end{equation}
where $H_\lambda$ is a new type of the Hamilton function with the parameter $\lambda$. Substituting \eqref{pc} into \eqref{HC2} and using $p=m\dot x$, we find
\begin{eqnarray}
H_\lambda(p,x)  &=& m\lambda^2\left[ \left( \frac{1}{\lambda^2} \int_0^p e^{-\frac{ \xi^2}{2m^2\lambda^2}} \frac{d \xi}{m} \right) \frac{p}{m} - m\lambda^2\left(   e^{-\frac{p^2}{2m^2\lambda^2}}   \right.\right.\nn\\
&&\left.\left.+\frac{p}{m^2\lambda^2} \int_0^p e^{-\frac{ \xi^2}{2m^2\lambda^2}} \frac{d \xi}{m} \right) \right] e^{-\frac{V(x)}{m\lambda^2}}\nn\\
 &=& -m\lambda^2  e^{-\frac{p^2}{2m^2\lambda^2}} e^{-\frac{V(x)}{m\lambda^2}}= -m\lambda^2  e^{-\frac{H_N}{m\lambda^2}} \;,
 \end{eqnarray}
which can be considered as the 1-parameter, namely $\lambda$, extended class of the Hamiltonian.
\\\\
Next, we show how to construct the multiplicative Hamiltonian.
It is assumed that the Hamiltonian takes an Ansatz form $H_\lambda=K(p)B(x)$ and it satisfies the Hamilton equation.
\begin{eqnarray}\label{HH}
 \dot x = \frac{\partial H_\lambda}{\partial p} \;\;\;\;,\;\;\;\dot p = -\frac{\partial H_\lambda}{\partial x}\;.
\end{eqnarray}
Using \eqref{HH}, the time derivative of the momentum $ p =m\dot x$ can be re-written in the form of
\begin{eqnarray} \label{HH12}
-\frac{\partial H_\lambda}{\partial x} &=& m \frac{d}{dt} \left(  \frac{\partial H_\lambda}{\partial p} \right)\nn \\
-\frac{1}{m}\frac{\partial H_\lambda}{\partial x} &=&  \dot p \frac{\partial^2 H_\lambda}{\partial p^2} + \frac{p}{m} \frac{\partial^2 H_\lambda}{\partial p \partial x}\;.
\end{eqnarray}
Inserting $H_\lambda$ into \eqref{HH12}, we obtain
\begin{equation}  \label{KB}
\frac{d^2 K}{d p^2} +    \frac{1}{m \dot p B}\frac{d B}{d x} \left(p \frac{d K}{d p} +  K \right)= 0\;.
\end{equation} 
To solve the differential equation, we define 
\begin{equation} \label{BA}
 \frac{1}{m \dot p B}\frac{d B}{d x}  \equiv W \;,
 \end{equation} 
where $W$ is a constant to be determined. Using the equation of motion \eqref{PE}, we find that the $B$ takes the form of
\begin{equation}
B(x) = \beta_1 e^{-mWV(x)}\;,
\end{equation}
where $\beta_1$ is a constant.
Substituting \eqref{BA} into \eqref{KB}, it is easily to see that
\begin{equation}
K(p) = \beta_2 e^{-\frac{Wp^2}{2}} \;,
\end{equation}
where $\beta_2$ is also a constant. Then the multiplicative Hamiltonian is given by 
$$  H_\lambda(p,x) = \kappa e^{-\frac{Wp^2}{2}} e^{-mAV(x)} \;,
$$
where  $\kappa=\beta_1\beta_2$ is a new constant. In order to determine the parameters $W$ and $\kappa$, we may proceed the same way as we did in the case of the Lagrangian by choosing $W=\frac{1}{m^2\lambda^2}$ and $\kappa=-m\lambda^2$ and considering the limit such that $\lambda$ approaches to infinity
\begin{eqnarray} \label{LH}
\lim_{\lambda \rightarrow \infty } H_\lambda &=& \lim_{\lambda \rightarrow \infty}\left[ -m\lambda^2 \left( 1  - \frac{p^2}{2m^2\lambda^2} + ...\right) \left( 1 - \frac{V(x)}{m\lambda^2} +... \right) \right]\nn\\
 &=&   \lim_{\lambda \rightarrow \infty}\left[ -m\lambda^2 \left( 1  - \frac{p^2}{2m^2\lambda^2}   - \frac{V(x)}{m\lambda^2} +... \right) \right]\nn\\
\lim_{\lambda \rightarrow \infty }\left( H_\lambda+ m\lambda^2 \right)&=&   \frac{p^2}{2m} + V(x)= H_N \;,
\end{eqnarray} 
which is the standard Hamiltonian in the additive form \eqref{HN}. Finally, the multiplicative Hamiltonian is
\begin{equation}\label{HC12} 
H_\lambda(p,x) = -m\lambda^2  e^{-\frac{H_N}{m\lambda^2}} \;,
\end{equation}
which is identical to the one obtained through Legendre transformation.

The last step is to show that the Hamiltonian in \eqref{HC12} yields the same equation of motion as that obtained in the Newtonian mechanics. In order to do this task, we substitute the Hamiltonian into \eqref{HH12} 
$$ \begin{array}{rcl}-  \frac{d V(x)}{d x} &=&   -\frac{p^2}{m^2\lambda^2} \frac{d V(x)}{d x} + \dot p \left( -\frac{p^2}{m^2\lambda^2} + 1 \right)
\\
-\frac{d V(x)}{d x}  \left( -\frac{p^2}{m^2\lambda^2} +1  \right) &=& \dot p  \left( -\frac{p^2}{m^2\lambda^2} + 1 \right)\;,\\
-\frac{d V(x)}{d x} &=&\dot p\;.
\end{array} 
$$
which is nothing but the equation of motion \eqref{PE}. 
\subsection{Infinite hierarchy of Lagrangians}\label{HL}
In the previous subsections, we managed to find the alternative forms of the Lagrangian and the Hamiltonian for the system with one degree of freedom. An extra-parameter $\lambda$ comes into the system naturally as the limit to the additional forms of the Lagrangian and the Hamiltonian. In this section, we will explore the role of the $\lambda$ in another aspect. In doing so, we start considering the expansion of the multiplicative form of Lagrangian in \eqref{SS1}
\begin{eqnarray}\label{dd}
\frac{L_\lambda}{m\lambda^2}
&=& \left(1 + \frac{T}{m\lambda^2} - \frac{1}{2! \cdot 3} \left( \frac{T}{m\lambda^2}\right)^2 
+ ... +  \frac{(-1)^{j-1}}{j! \cdot 2j-1}\left(\frac{T}{m\lambda^2} \right)^j + ...\right)\nn\\
&& \left(1 - \frac{V}{m\lambda^2} + \frac{1}{2!}\left(\frac{V}{m\lambda^2}\right)^2 
+ ... + \frac{(-1)^j}{j!}\left(\frac{V}{m\lambda^2}\right)^j + ...\right) \nn\\
&=& 1 +   \frac{1}{1!m\lambda^2}[T-V] - \frac{1}{2! (m\lambda^2)^2}\left[ \frac{T^2}{3} + 2TV  -V^2\right] \nn\\
&&+ \frac{1}{3!(m\lambda^2)^3}\left[ \frac{T^3}{5} +T^2V + 3TV^2 - V^3\right]  +...\nn\\
&&+ \frac{(-1)^{j-1}}{j!(m\lambda^2)^j}\left[ \frac{T^j}{2j-1} + \frac{jT^{j-1}V}{2j-3} + \frac{j(j-1)T^{j-2}V^2}{2!(2j-5)} + ...+ jTV^{j-1} - V^j\right]\nn\\
&& +.... \;\; ,
\end{eqnarray}
where the kinetic energy is a function of velocity variable: $T = T(\dot x)$. Collecting the coefficients in the expansion, we obtain
\begin{subequations}\label{hi2}
\begin{eqnarray}
\mathcal{O}\left(\frac{1}{1!(m\lambda^2)^1}\right):\mathcal{L}_1&\equiv&T-V\;.\\
\mathcal{O}\left(\frac{1}{2!(m\lambda^2)^2}\right):\mathcal{L}_2&\equiv&\frac{T^2}{3} + 2TV  -V^2\;.\\
\mathcal{O}\left(\frac{1}{3!(m\lambda^2)^3}\right):\mathcal{L}_3&\equiv&\frac{T^3}{5} +T^2V + 3TV^2 - V^3\;.\\
&.&\nn\\
&.&\nn\\
&.&\nn\\
\mathcal{O}\left(\frac{1}{j!(m\lambda^2)^j}\right):\mathcal{L}_j&\equiv&\frac{T^j}{2j-1} + \frac{jT^{j-1}V}{2j-3} + \frac{j(j-1)T^{j-2}V^2}{2!(2j-5)} \nn\\
&&+ ...+ jTV^{j-1} - V^j\;.\\
&.&\nn\\
&.&\nn\\
&.&\nn
\end{eqnarray}
\end{subequations}
which form a set of infinite number of Lagrange functions: $\{\mathcal{L}_1,\mathcal{L}_2,\mathcal{L}_3,...,\mathcal{L}_j,...\}$, called an infinite hierarchy of Lagrangians. Furthermore, we find that
\begin{eqnarray}
\mathcal{L}_{j-1}&=&\frac{1}{j}\frac{\partial \mathcal{L}_j}{\partial V}\;,\\
\mathcal{L}_{1}&=&\frac{1}{j!}\frac{\partial^{j-1} \mathcal{L}_j}{\partial V^{j-1}}\;,
\end{eqnarray}
if we treat the potential energy as a variable.
With \eqref{hi2}, \eqref{dd} can be written in a compact form
\begin{equation}
L_\lambda=\sum_{j=0}^\infty\frac{1}{j!}\left(\frac{-1}{m\lambda^2}\right)^{j-1}\mathcal{L}_j\;,
\end{equation}
where 
\begin{equation}
\mathcal{L}_j=\sum_{k=0}^{j}\left[  \frac{j!T^{j-k}V^k}{(j-k)!k!(2j-(2k+1))} \right]\;.
\end{equation}
Interestingly, Lagrangians in the hierarchy produce the same equation of motion \eqref{NE}. To see this, we compute
$$\begin{array}{rcl}
\Rightarrow \frac{\partial \mathcal{L}_j}{\partial x} &=& \sum_{k=1}^{j}\left[  \frac{j!T^{j-k}V^{k-1}}{(j-k)!(k-1)!(2j-(2k+1))} \right]\frac{d V}{d x}    \hspace{2cm} \\
&=& \sum_{k=0}^{j-1}\left[  \frac{j!T^{j-k-1}V^{k}}{(j-k-1))!k!(2j-(2k+3))} \right]\frac{d V}{d x} \\
&=& \left( \sum_{k=0}^{j-2}\left[  \frac{j!T^{j-k-1}V^{k}}{(j-k-1))!k!(2j-(2k+3))} \right] +\frac{j!V^{j-1}}{(j-1)!(-1)}  \right)\frac{d V}{d x}\;.
\end{array}$$
$$\begin{array}{rcl}
\Rightarrow\frac{\partial \mathcal{L}_j}{\partial \dot x} &=&  \sum_{k=0}^{j-1}\left[  \frac{j!T^{j-k-1}V^k}{(j-k-1)!k!(2j-(2k+1))} \right]m\dot x\;.
 \\\\
\Rightarrow\dot x \frac{\partial^2 \mathcal{L}_j}{\partial x \partial \dot x} &=&  \sum_{k=1}^{j-1}\left[  \frac{j!T^{j-k-1}V^{k-1}}{(j-k-1)!(k-1)!(2j-(2k+1))} \right]m\dot x^2 \frac{d V}{d x}  \hspace{1cm}  \\

&=&    \sum_{k=1}^{j-1}\left[  \frac{j!T^{j-k}V^{k-1}}{(j-k-1)!(k-1)!(2j-(2k+1))} \right]2 \frac{d V}{d x}  \hspace{2cm}  \\ 
&=& \sum_{k=0}^{j-2}\left[  \frac{2j!T^{j-k-1}V^{k}}{(j-k-2)!k!(2j-(2k+3))} \right] \frac{d V}{dl x}\;.\\
\\
\end{array}$$
$$\begin{array}{rcl}
\Rightarrow\ddot x \frac{\partial^2 \mathcal{L}_j}{\partial^2 \dot x} &=& \ddot x \left(\sum_{k=0}^{j-1}\left[  \frac{j!T^{j-k-1}V^k}{(j-k-1)!k!(2j-(2k+1))} \right]m\right. \\
&&+  \left.\sum_{k=0}^{j-2}\left[  \frac{j!T^{j-k-2}V^k}{(j-k-2)!k!(2j-(2k+1))} \right]m^2\dot x^2 \right) \\
&=&  \sum_{k=0}^{j-1}\left[  \frac{j!T^{j-k-1}V^k}{(j-k-1)!k!(2j-(2k+1))} \right]m\ddot x \\
&&+  \sum_{k=0}^{j-2}\left[  \frac{2j!T^{j-k-1}V^k}{(j-k-2)!k!(2j-(2k+1))} \right]m \ddot x \\
 &=&  \left( \sum_{k=0}^{j-2}\left[  \frac{j!T^{j-k-1}V^k}{(j-k-1)!k!}\left( \frac{1+2(j-k-1)}{(2j-(2k+1))} \right) \right] + \frac{j!V^{j-1}}{(j-1)!}\right)m \ddot x \\
 &=& \left( \sum_{k=0}^{j-2}\left[  \frac{j!T^{j-k-1}V^k}{(j-k-1)!k!} \right] + \frac{j!V^{j-1}}{(j-1)!}\right)m \ddot x\;.
\end{array}$$
We insert above results into the Euler-Lagrange equation
$$\frac{\partial \mathcal{L}_j}{\partial x} - \dot x \frac{\partial^2 \mathcal{L}_j}{\partial x \partial \dot x} - \ddot x \frac{\partial^2 \mathcal{L}_j}{\partial^2 \dot x}=0\;, $$
which yields the equation of motion \eqref{NE}.
 \\\\
According to the structure of the infinite hierarchy of Lagrangians, we may consider the multiplicative form of the Lagrangian as a generating function for Lagrange polynomial\footnote{Of cause, we did not mean to Lagrange interpolating polynomial, but rather polynomial in terms of kinetic and potential energies.} in \eqref{hi2} and the parameter $\lambda$ plays as a mathematical tool to go from the multiplicative form to the additive form of the Lagrangian.
\\
\\
Next, we will look for the Hamiltonian corresponding to each Lagrangian $\mathcal{L}_j$ in the hierarchy. First, we may start by working on Legendre transformation
\begin{eqnarray}\label{LTS}
\mathcal{H}_j &=& \dot x \frac{\partial \mathcal{ L}_j}{\partial \dot x} - \mathcal{L}_j \nn \\
&=&    \sum_{k=0}^{j-1}\left[  \frac{2j!T^{j-k}V^k}{(j-k-1)!k!(2j-(2k+1))} \right]  -   \sum_{k=0}^{j}\left[  \frac{j!T^{j-k}V^k}{(j-k)!k!(2j-(2k+1))} \right]  \nn \\
&=& \sum_{k=0}^{j-1}\left[  \frac{j!T^{j-k}V^k}{(j-k)!k!}\left( \frac{2(j-k) -1}{(2j-(2k+1))}  \right) - \frac{V^j}{(-1)}\right]   \nn\\
&=&    \sum_{k=0}^{j-1}\left[  \frac{j!T^{j-k}V^k}{(j-k)!k!} +V^j \right] 
=   \sum_{k=0}^{j}\left[  \frac{j!T^{j-k}V^k}{(j-k)!k!} \right]  \nn\\
&=&   (T+V)^j=H_N^j\;,
\end{eqnarray}
where the kinetic energy is expressed in terms of the momentum variable: $T = T(p)$.
Second, we may perform the expansion on the Multiplicative Hamiltonian \begin{eqnarray}
H_\lambda &=&  -m\lambda^2  e^{-\frac{T+V}{m\lambda^2}}   \nn\\
&=&  \sum_{j=0}^{\infty} \frac{1}{j!}\left(\frac{-1}{m\lambda^2}\right)^{j-1}   (T+V)^j  \nn\\
&=&  \sum_{j=0}^{\infty}   \frac{1}{j!}\left(\frac{-1}{m\lambda^2}\right)^{j-1}   \mathcal{H}_j \;.
\end{eqnarray}
Then we now also obtain a hierarchy of the Hamiltonians: $\{\mathcal{H}_1,\mathcal{H}_2,\mathcal{H}_3,...,\mathcal{H}_j,... \}=\{H_N,H_N^2,H_N^3,...,H_N^j,... \}$ which can be transformed to Lagrangian hierarchy \eqref{hi2} through the Legendre transformation \eqref{LTS}.
\\\\
\emph{Note}: For a harmonic oscillator, the hierarchy of the Hamiltonians can be obtained through the existence of the spatial Lax matrix $\boldsymbol{L}$ given by \cite{Babe}
\begin{equation}
\boldsymbol{L}=
\left( \begin{array}{cc}
p & \omega x \\
\omega x &  -p
\end{array} \right)\;,
\end{equation}
where the mass is set to be unity. The conserved quantities or Hamiltonians are $\mbox{Tr}(\boldsymbol{L}^{2l})=2(2\mathcal{H})^l$, where $2\mathcal{H}=p^2+\omega^2x^2$, and $\mbox{Tr}(\boldsymbol{L}^{2l+1})=0$.
%
%
%
%
%
%
%
%
%
%
%
\section{Relativistic case}\label{RL}
In this section, we extend the idea to the case of the relativistic case with a system of one degree of freedom for searching the multiplicative forms of Lagrangian and Hamiltonian. The equation of motion of the system is given by
\begin{eqnarray} 
 m\ddot x \gamma^3 = -\frac{d V(x)}{dx}\;, \label{EQ}
\end{eqnarray}
where $m$ is defined as the rest mass and
\[
\gamma=\frac{1}{\sqrt{1-\frac{\dot x^2}{c^2}}}\;.
\]
The parameter $c$ is the speed of light. The standard additive form of relativistic Lagrangian and relativistic Hamiltonian are given by 
\begin{eqnarray}
L_c(\dot x,x)  &=&  - \frac{mc^2}{\gamma} - V(x)\;,  \label{LC}\\
H_c(p,x) &=& \gamma mc^2+V(x)= mc^2\sqrt{1+\left(\frac{p}{mc}\right)^2} + V(x)\;, \label{HC}
\end{eqnarray}
where $p=\gamma m\dot x$ is the relativistic momentum.

\subsection{Multiplicative relativistic Lagrangian}

In this section, we will proceed the same technique as shown in section \ref{ML} to solve the multiplicative form of the relativistic Lagrangian for the system in \eqref{EQ}. We start to introduce the Ansatz form of the Lagrangian as
$L  =  M(\dot x) N(x)$ , where $M(\dot x)$ and $N(x)$ will be determined. We substitute the Lagrangian into the Euler-Lagrangian equation
\begin{eqnarray}
\frac{\partial L}{\partial x} - \frac{d}{dt}\left(   \frac{\partial L}{\partial \dot x}\right) = 0\;.
\end{eqnarray}
Then we obtain
\begin{eqnarray}
\frac{d^2 M}{d \dot x^2} +\left( \dot x \frac{d M}{d \dot x}  -  M\right) \frac{1}{\ddot x N}\frac{d N}{dx} &=& 0\;, \label{22}
\end{eqnarray}
and we impose the equation of motion on \eqref{22} yielding
\begin{eqnarray}
\frac{d^2 M}{d \dot x^2} +\left( \dot x \frac{d M}{d \dot x}  -  M\right)\gamma^3 C_1 &=& 0\;,\label{MM}
\end{eqnarray}
where
\begin{eqnarray}
\frac{1}{N} \frac{d N}{d x} = -\frac{C_1}{m}\frac{d V}{d x}\;,\label{NN}
\end{eqnarray}
and $C_1$ is to be identified.
Solving \eqref{MM} and \eqref{NN}, we obtain $M$ and $N$ in the form
\begin{eqnarray} 
M(\dot x)  &=& -C_3 \left(    e^{-C_1c^2 \gamma}  + C_1 \dot x\int_0^{\dot x}   \gamma_v^3 e^{-C_1c^2\gamma_v}  d v\right) \;,\;\mbox{and}\;\;\gamma_v=\frac{1}{\sqrt{1-\frac{v^2}{c^2}}}\;, \label{M1}\\
N(x) &=& C_2 e^{-\frac{C_1V(x)}{m}}\;,  \label{N1}
\end{eqnarray}
where $C_2$ and $C_3$ are constants to be determined. Using \eqref{M1} and \eqref{N1}, the multiplicative Lagrangian becomes
$$\begin{array}{rcl} 
L  (x,\dot x)
&=& -C_2C_3 \left(   e^{-C_1c^2 \gamma}  + C_1 \dot x\int_0^{\dot x}   \gamma_v^3 e^{-C_1c^2\gamma_v}  d v\right)e^{-\frac{C_1V(x)}{m}}\;.
\end{array}$$ 
We choose  $C_2C_3 = -m\lambda^2$ and $C_1 = \frac{1}{\lambda^2}$, where $\lambda$ is again in the velocity unit. The multiplicative relativistic Lagrangian takes the form
\begin{equation}  
L_{\lambda,c}(x,\dot x)   =   m\lambda^2 \left( e^{- \frac{\gamma c^2}{\lambda^2} } + \frac{\dot x}{\lambda ^2} \int_0^{\dot x}  \gamma_v^3  e^{- \frac{\gamma_v c^2}{\lambda^2}}    d  v\right) e^ {-\frac{V(x)}{m\lambda^2}}\;,\label{LLC}
\end{equation}
which depends on two parameters, i.e., $\lambda$ and $c$. To see the role of these parameters, we consider the following situations.
\\\\
\textbf{Non-multiplicative limit}: We find that if $\lambda$ is very large the multiplicative relativistic Lagrangian is reduced to the standard relativistic Lagrangian $\eqref{LC}$
\begin{eqnarray}\label{NML}
 \lim_{\lambda \rightarrow \infty}   L_{\lambda,c}(x,\dot x) 
&=&m \lambda^2 \left( 1-  \frac{\gamma c^2}{\lambda^2} + \frac{\dot x}{\lambda^2} \int_0^{\dot x} \gamma_v^3 d v - \frac{V(x)}{m\lambda^2}  \right) \nn\\
&=& m\lambda^2 - mc^2\left( \frac{1- \frac{\dot x^2}{c^2}}{\sqrt{1-\frac{\dot x^2}{c^2}}}  \right)  - V(x)\nn\\
&=& m\lambda^2 + L_c(\dot x,x)\;.  
\end{eqnarray}
\textbf{Non-relativistic limit}: If we take the limit: $\dot x << c$ the multiplicative relativistic Lagrangian becomes the multiplicative non-relativistic Lagrangian 
\begin{eqnarray}\label{NRL}
\lim_{\dot x<< c}  L_{\lambda,c}(x,\dot x) &=& \lim_{\dot x<< c}  m\lambda^2 \left[  e^{- \frac{c^2}{\lambda^2} {\left( 1 + \frac{\dot x^2}{2c^2} + ...\right)}} + \frac{\dot x}{\lambda ^2} \int_0^{\dot x}     \left( 1 + \frac{3v^2}{2c^2} + ... \right)  \right.\nn\\
&&\;\;\;\;\;\;\;\;\;\;\;\;\;\;\;\;\;\;\;\left.\times e^{- \frac{c^2}{\lambda^2} {\left( 1 + \frac{v^2}{2c^2} + ...\right)}}    d v\right] e^ {-\frac{V(x)}{m\lambda^2}} \nn\\
&=&    m\lambda^2 e^{-\frac{c^2}{\lambda ^2}} \left(  e^{- \frac{\dot x^2}{2\lambda^2} } + \frac{\dot x}{\lambda^2} \int_0^{\dot x}       e^{  - \frac{v^2}{2\lambda^2} }  d  v\right) e^ {-\frac{V(x)}{m\lambda^2}} \nn\\
&=& e^{-\frac{c^2}{\lambda ^2}} L_\lambda(x,\dot x) \;,
\end{eqnarray}
which the factor $e^{-\frac{c^2}{\lambda ^2}}$ makes no effect on the equation of motion.
\\\\
When taking both limits in either sequences, it leads to the same result
\begin{eqnarray}
 \lim_{\substack{\dot x<< c\\ \lambda \rightarrow \infty}}   L_{\lambda,c}(x,\dot x) = m\lambda^2-mc^2 + L_N(\dot x,x)\;,
\end{eqnarray}
which is the standard non-relativistic Lagrangian. 
\\\\
The last task is to demonstrate that the multiplicative Lagrangian \eqref{LLC} gives the correct equation of motion. We find that
$$\begin{array}{rcl}
\frac{\partial L_{\lambda,c}}{\partial \dot x} &=& m e^{-\frac{V}{m\lambda^2}}    \int_0^{\dot x}     \left( 1 - \frac{v^2}{c^2} \right)^{-\frac{3}{2}}  e^{- \frac{c^2}{\lambda^2} {\left( 1 - \frac{v^2}{c^2} \right)}^{-\frac{1}{2}}}    d v \;.\\
\frac{d}{dt}\left( \frac{\partial L_{\lambda,c}}{\partial \dot x} \right) &=& m \ddot x e^{-\frac{V}{m\lambda^2}}   \left( 1 - \frac{\dot x^2}{c^2} \right)^{-\frac{3}{2}}  e^{- \frac{c^2}{\lambda^2} {\left( 1 - \frac{\dot x^2}{c^2} \right)}^{-\frac{1}{2}}} \\
&& 
-\frac{\dot x  e^{-\frac{V}{m\lambda^2}}}{\lambda^2}   \left[  \int_0^{\dot x} \left( 1 - \frac{v^2}{c^2} \right)^{-\frac{3}{2}}  e^{- \frac{c^2}{\lambda^2} {\left( 1 - \frac{v^2}{c^2} \right)}^{-\frac{1}{2}}} d v\right]      \frac{d V}{dx}\;. \\
\frac{\partial L_{\lambda,c}}{\partial x} &=&- e^{-\frac{V}{m\lambda^2}}   \left[ e^{- \frac{c^2}{\lambda^2} {\left( 1 - \frac{\dot x^2}{c^2} \right)}^{-\frac{1}{2}}} + \frac{\dot x}{\lambda^2} \int_0^{\dot x}    \left( 1 - \frac{v^2}{c^2} \right)^{-\frac{3}{2}}  e^{- \frac{c^2}{\lambda^2} {\left( 1 - \frac{v^2}{c^2} \right)}^{-\frac{1}{2}}}   d v \right] \frac{dV}{d x}\;.
\end{array}$$
We then substitute into the Euler-Lagrange equation
$$\begin{array}{rcl} \frac{\partial L_{\lambda,c}}{\partial x} - \frac{d}{dt}\left( \frac{\partial L_{\lambda,c}}{\partial \dot x} \right) &=& 0\;, \\
\end{array}$$
yielding 
 \eqref{EQ}.


\subsection{Multiplicative relativistic Hamiltonian}
In the previous subsection, we successfully constructed the multiplicative relativistic Lagrangian. In this section, we will compute the corresponding multiplicative relativistic Hamiltonian. We may start to perform the naive Legendre transformation
\begin{eqnarray}\label{HMC}
H_{\lambda,c} &=& \dot x\frac{\partial L_{\lambda,c}}{\partial \dot x} - L_{\lambda,c} \nn \\
&=&  \dot x \left(    \int_0^{\dot x}   \gamma_v^3  e^{- \frac{\gamma_v c^2}{\lambda^2}}    d  v\right)  e^{-\frac{V(x)}{m\lambda^2}} \nn - m\lambda^2 \left(  e^{- \frac{\gamma c^2}{\lambda^2} } + \frac{\dot x}{\lambda ^2} \int_0^{\dot x}  \gamma_v^3  e^{- \frac{\gamma_v c^2}{\lambda^2}}    dv\right) e^ {-\frac{V(x)}{m\lambda^2}} \nn \\
&=& -m\lambda^2 e^{- \frac{\gamma c^2}{\lambda^2} } e^ {-\frac{V(x)}{m\lambda^2}}
=-m\lambda^2 e^{- \frac{H_c}{m\lambda^2} }\;,  
\end{eqnarray}
which is in the multiplicative form. 
\\\\
Next we will directly construct the multiplicative relativistic Hamiltonian. Here we take the Ansatz form of the Hamiltonian $H_{\lambda,c}(p,x) = P(p)Q(x)$ and use the Hamilton's equation
\begin{eqnarray}\
 \dot x = \frac{\partial H_{\lambda,c}}{\partial p} \;\;\;\;,\;\;\;\dot p = -\frac{\partial H_{\lambda,c}}{\partial x}\;.
\end{eqnarray}
Equating these two equations, we obtian
\begin{eqnarray} \label{HH2}
-\frac{\partial H_{\lambda,c}}{\partial x} &=& \frac{d}{dt} \left(  \gamma m  \frac{\partial H_{\lambda,c}}{\partial p} \right) \label{42}\nn \\
-\frac{1}{m}\frac{\partial H_{\lambda,c}}{\partial x} &=&  \gamma \left( \dot p \frac{\partial^2 H_{\lambda,c}}{\partial p^2} + \frac{p}{\gamma m} \frac{\partial^2 H_{\lambda,c}}{\partial p \partial x}\right) + \frac{\partial H_{\lambda,c}}{\partial p} \frac{p \dot p}{\gamma m^2c^2}\;. \label{HC1}
\end{eqnarray}
Inserting $H_{\lambda,c}$ into \eqref{HC1}, we obtain 
\begin{eqnarray}  
  \gamma \frac{d^2 P}{d p^2} +    \frac{1}{ m \dot p Q}\frac{d Q}{d x} \left(p \frac{d P}{d p} +  P \right) + \frac{p}{\gamma m^2c^2} \frac{d P}{d p}&=& 0\;.
\end{eqnarray} 
We now set
\begin{equation}
\frac{1}{m\dot p Q}\frac{d Q}{d x} = C_4\Rightarrow \frac{1}{ Q}\frac{d Q}{d x} = -m\frac{dV}{dx}C_4\;,\label{C4}
\end{equation} 
where $\dot p = -d V/dx$ has been used and $C_4$ is a constant to be determined. We then have 
\begin{eqnarray}  \gamma \frac{d^2 P}{d p^2} + \frac{p}{\gamma m^2c^2} \frac{d P}{d p} +  C_4 \left(p \frac{d P}{d p} +  P \right)&=& 0\;. \label{PP}
\end{eqnarray}
Solving \eqref{C4} and \eqref{PP}, we obtain functions $P$ and $Q$ leading to the Hamiltonian
$$ H_{\lambda,c} = C_5 e^{-C_4 m^2c^2\gamma} e^{-C_4mV(x)}\;,$$
where $C_5$ is a constant to be also determined. We now choose $C_4 = \frac{1}{m^2\lambda^2}$ and  $C_5 = -m\lambda^2$ leading to the multiplicative relativistic Hamiltonian in form
\begin{eqnarray} H_{\lambda,c}(p,x)  = -m\lambda^2 e^{-\frac{\gamma c^2}{\lambda^2}}e^{-\frac{V(x)}{m\lambda^2}}\;,\label{RMH}
\end{eqnarray}
which is identical to \eqref{HMC}.
\\
\\
\textbf{Non-multiplicative limit}: For the case of vary large $\lambda$, we have
\begin{eqnarray}\label{HNML}
\lim_{\lambda \rightarrow \infty} H_{\lambda,c} &=& - \lim_{\lambda \rightarrow \infty}\left[ m\lambda^2\left(  1- \frac{\gamma c^2}{\lambda^2} +  \frac{\gamma^2 c^4}{2!\lambda^4} + ...  \right)\left(  1 - \frac{V}{m \lambda^2} + \frac{V^2}{2!m^2\lambda^4} + ...  \right)\right] \nn   \\
&=&    - m\lambda^2  + H_c\;,
\end{eqnarray}
where $H_c$ is the standard relativistic Hamiltonian.
\\
\\
\textbf{Non-relativistic limit}: The multiplicative Hamiltonian  \eqref{RMH} can be reduced to the form by consider the non-relativistic limit: $p<<mc$
\begin{eqnarray} \label{HNRL}
\lim_{p << mc} H_{\lambda,c} &=&- \lim_{p << mc} \left[ m_0\lambda^2 e^{-\frac{c^2}{\lambda^2}\left(  1 +   \frac{p^2}{2m^2 c^2} + ... \right)}e^{-\frac{V}{m\lambda^2}} \right]  \nn \\
&=&e^{-\frac{c^2}{\lambda^2}} H_\lambda\;, 
\end{eqnarray}
where $H_\lambda$ is the multiplicative non-relativistic Hamiltonian.
\\
\\
Again when taking both limits in either sequences, it leads to the same result
\begin{eqnarray}
 \lim_{\substack{\dot x<< c\\ \lambda \rightarrow \infty}}   H_{\lambda,c}(p,x) = -m\lambda^2+mc^2 + H_N(p,x)\;,
\end{eqnarray}
which is the standard non-relativistic Hamiltonian.
\\
\\
The last step, we will show that the multiplicative relativistic Hamiltonian gives the correct relativistic equation of motion. Substituting Hamiltonian \eqref{RMH} in \eqref{HH2}, we obtain
\begin{eqnarray} 
-e^{-\frac{\gamma c^2}{\lambda^2}} e^{-\frac{V}{m\lambda^2}} \frac{d V}{d x}  
&=& e^{-\frac{\gamma c^2}{\lambda^2}} e^{-\frac{V}{m\lambda^2}}   \left[ \left( 1 -\frac{p^2}{\gamma m^2\lambda^2} \right)\dot p   - \frac{p^2}{\gamma m^2\lambda^2}  \frac{d V}{dx}\right] \nn  \\
0&=&  e^{-\frac{\gamma c^2}{\lambda^2}} e^{-\frac{V}{m\lambda^2}}   \left[ \left( 1 -\frac{p^2}{\gamma m^2\lambda^2} \right)\left(\dot p +  \frac{dV}{d x}\right)\right] \nn\\
\dot p &=& - \frac{dV}{dx}\;,
\end{eqnarray}
which is the desired equation of motion \eqref{EQ}.

\subsection{Infinite hierarchy of relativistic Lagrangian}
In previous subsections, we obtained the multiplicative relativistic Lagrangian and Hamiltonian that yield the same relativistic equation of motion as those for the case of the standard Lagrangian and Hamiltonian. The two successive limits have been performed, i.e., on the parameter $\lambda$ and on the parameter $c$.
\\\\
In this section, we consider alternative limit on the parameter $\lambda$ called the perturbative limit. We begin to consider the expansion multiplicative relativistic Lagrangian with respect to the parameter $\lambda$ 
\begin{eqnarray}    
L_{\lambda.c} &=& m\lambda^ 2 \left[ \sum_{j=0}^{\infty}\frac{1}{j!}\left(-\frac{ c^2}{\lambda^2}\right)^j \gamma^j  + \frac{\dot x}{\lambda^2}\int_0^{\dot x} \sum_{j=0}^{\infty}\frac{1}{j!}\left(-\frac{ c^2}{\lambda^2}\right)^j \gamma_v^{j+3}  dv  \right]e^ {-\frac{V}{m\lambda^2}}  \nn \\
&=&  m\lambda^ 2 \left[ 1 + \sum_{j=1}^{\infty} \frac{1}{j!}\left(-\frac{ c^2}{\lambda^2}\right)^j \gamma^j   - \frac{\dot x}{c^2}\int_0^{\dot x} \sum_{j=1}^{\infty}\frac{1}{(j-1)!}\left(-\frac{ c^2}{\lambda^2}\right)^{j} \gamma_v^{j+2}   dv  \right]e^ {-\frac{V}{m\lambda^2}}  \nn \\
&=& m\lambda^2\left[ 1 + \sum_{j=1}^{\infty}\frac{1}{j!} \left(- \frac{c^2}{\lambda^2} \right)^j \left(  \gamma^j - \frac{j\dot x}{c^2} \int_0^{\dot x} \gamma_v^{j+2} d v \right) \right] e^ {-\frac{V}{m\lambda^2}}\;. \label{LC1} 
\end{eqnarray}
We now introduce the functions $ \mathcal{P}_j =   \gamma^j - \frac{j\dot x}{c^2} \int_0^{\dot x}\gamma_v^{j+2} dv $, i.e.,
\begin{eqnarray}
\mathcal{P}_1 
&=& \frac{1}{\gamma} \;,\nn \\
\mathcal{P}_2 
&=& 1 - \frac{\dot x}{c}\ln{\left(\frac{\dot x\gamma^2}{c}\right)} \;,\nn \\
\mathcal{P}_3 
&=& \gamma\left(1  -\frac{2\dot x^2}{c^2} \right)\;,\nn \\
\mathcal{P}_4 
&=&\gamma^2\left(1-\frac{3\dot x^2}{2c^2} \right) -\frac{3\dot x}{2c}\ln{\left(\frac{\dot x\gamma^2}{c}\right)}\;, \nn \\
&.& \nn\\
&.& \nn\\
&.& \nn
\end{eqnarray}
Therefore, the Lagrangian \eqref{LC1} can be expanded in terms of functions $\mathcal{P}_j$
\begin{eqnarray}
L_{\lambda.c} 
&=&  m\lambda^2 \left[  1 -\frac{c^2 }{\lambda^2}\mathcal{P}_1 + \frac{c^4}{2!\lambda^4}\mathcal{P}_2  -  \frac{c^6}{3!\lambda^6}\mathcal{P}_3+...\right] \left[ 1 - \frac{V}{m\lambda^2} + \frac{V^2}{2!(m\lambda^2)^2}\right.\nn\\
&&\left. - \frac{V^3}{3!(m\lambda^2)^3} + ...\right] \nn \\
&=& m\lambda^2 - \left[ mc^2\mathcal{P}_1  + V\right] + \frac{1}{2! m\lambda^2}\left[ (mc^2)^2\mathcal{P}_2 + 2mc^2\mathcal{P}_1V + V^2\right] \nn \\
&&- \frac{1}{3!(m\lambda^2)^2}\left[ (mc^2)^3\mathcal{P}_3  + 3 (mc^2)^2\mathcal{P}_2V + 3(mc^2)\mathcal{P}_1V^2 + V^3 \right] + .... \nn \\
&&- \frac{1}{j!}\left(\frac{-1}{m\lambda^2}\right)^{j-1}\left[ (mc^2)^j\mathcal{P}_j + j (mc^2)^{j-1}\mathcal{P}_{j-1}V \right.\nn\\
&&\left.+ j(j-1) (mc^2)^{j-2}\mathcal{P}_{j-2}V^2 + ... +V^j\right]+....\; \nn \\
&=& \sum_{j=0}^{\infty}  \frac{1}{j!}\left(\frac{-1}{m\lambda^2}\right)^{j-1} \mathcal{L}_{j,c}\;, \label{lc}
\end{eqnarray}
where 
\begin{equation}
\mathcal{L}_{j,c} = -  \sum_{k=0}^{j}\left[ \frac{j!}{(j-k)!k!}(mc^2)^{j-k}\mathcal{P}_{j-k}V^k\right] \;.
 \end{equation}
Equation \eqref{lc} suggests that there exists an infinite set $\{\mathcal{L}_{1,c},\mathcal{L}_{2,c},\mathcal{L}_{3,c},...,\mathcal{L}_{j,c},...\}$, called an infinite hierarchy of relativistic Lagrangians, such that
\begin{subequations}\label{hi}
\begin{eqnarray}
\mathcal{L}_{1,c}&\equiv& -\mathcal{P}_1m_0c^2 - V\\
\mathcal{L}_{2,c}&\equiv& - \mathcal{P}_{2}(m_0c^2)^2 -2\mathcal{P}_{1}m_0c^2 V - V^2 \\
\mathcal{L}_{3,c}&\equiv&-\mathcal{P}_{3}(m_0c^2)^3 -3\mathcal{P}_{2}(m_0c^2)^2V -3\mathcal{P}_{1}m_0c^2V^2 - V^3\\
&.&\nn\\
&.&\nn\\
&.&\nn\\
\mathcal{L}_j,c&\equiv& -\mathcal{P}_j(m_0c^2)^j - j\mathcal{P}_{j-1} (m_0c^2)^{j-1}V - j(j-1)\mathcal{P}_{j-2} (m_0c^2)^{j-2}V^2 - ...\nn \\
&&- j\mathcal{P}_1 V^{j-1}-V^j\\
&.&\nn\\
&.&\nn\\
&.&\nn
\end{eqnarray}
\end{subequations}
Furthermore, we find the relations
\begin{eqnarray}
\frac{\partial \mathcal{L}_{j,c}}{\partial V} &=& j \mathcal{L}_{j-1,c}\;, \\
\frac{\partial^{j-1} \mathcal{L}_{j,c}}{\partial V^{j-1}} &=& j! \mathcal{L}_{1,c}\;.
\end{eqnarray}
An interesting point is that all Lagrangians in \eqref{hi} produce the same relativistic equations of motion \eqref{EQ}. In order to prove the statement, we compute
$$\begin{array}{rcl}
\frac{\partial\mathcal{L}_{j,c}}{\partial \dot x} &=&  \sum_{k=0}^{j-1}\left[\frac{j!}{(j-k)!k!}\left( \frac{j-k}{c^2} \right) \int_0^{\dot x} \gamma_v^{j-k+2} d v(mc^2)^{j-k}V^k\right] \;,\\
\ddot x\frac{\partial^2 \mathcal{L}_{j,c}}{\partial \dot x^2} &=&    \sum_{k=0}^{j-1}\left[ \frac{j!}{(j-k)!k!}\left( \frac{j-k}{c^2} \right) \gamma_v^{j-k+2} (mc^2)^{j-k}V^k\right]\ddot x \;,\\
\dot x \frac{\partial^2 \mathcal{L}_{j,c}}{\partial x \partial \dot x}  &=&      \sum_{k=1}^{j-1}\left[ \frac{j!}{(j-k)!(j-1)!}\left( \frac{j-k}{c^2} \right) \int_0^{\dot x} \gamma_v^{j-k+2} dv(mc^2)^{j-k}V^{k-1}\right] \dot x \frac{d V}{d x} \;,\\
\frac{\partial \mathcal{L}_{j,c}}{\partial x} &= & -\sum_{k=1}^{j}\left[ \frac{j!}{(j-k)!(k-1)!}(mc^2)^{j-k}\mathcal{P}_{j-k}V^{k-1}\right] \frac{d V}{d x} \;.\\
\end{array}$$ 
Substituting into the Euler-Lagrange equation
\begin{eqnarray}
 \frac{\partial  \mathcal{L}_{j,c}}{\partial x}  -  \dot x\frac{\partial^2  \mathcal{L}_{j,c}}{\partial x \partial \dot x} &=& \ddot x \frac{\partial^2   \mathcal{L}_{j,c}}{\partial \dot x^2} \;,
\end{eqnarray}
which gives the relativistic equation of motion \eqref{EQ}.
\\
\\
By working on Legendre transformation, we find the Hamiltonian that corresponds to the Lagrangian $ \mathcal{L}_{j,c}$
\begin{eqnarray} \mathcal{H}_{j,c} &=& \dot x \frac{\partial \mathcal{L}_{j,c}}{\partial \dot x} - \mathcal{L}_{j,c} \nn \\
&=& -    \sum_{k=0}^{j-1}\left[   \frac{j!}{(j-k)!k!}(mc^2)^{j-k} \left( -\frac{(j-k)\dot x}{c^2} \int_0^{\dot x} \gamma_v^{j-k+2} dv   - \mathcal{P}_{j-k}\right) V^k \right] -V^k \nn \\
&=&   \sum_{k=0}^{j}  \frac{j!}{(j-k)!k!}(\gamma mc^2)^{j-k}V^k  = \left( \gamma mc^2 + V\right)^j = H_c^j\;.\label{Leg}
\end{eqnarray}
Equation \eqref{Leg} suggests that by performing the Legendre transformation on each Lagrangian in the hierarchy we can construct also the infinite hierarchy of Hamiltonians: $\{\mathcal{H}_{1,c},\mathcal{H}_{2,c},...,\mathcal{H}_{j,c},...\}=\{ H_c,H_c^2,....,H_c^j,...\}$. 
Alternatively, we can directly expand the multiplicative relativistic Hamiltonian
\begin{eqnarray}
H_{\lambda,c}  &=& -m\lambda^2 \left( 1-\frac{\gamma c^2}{\lambda^2}+\frac{1}{2}\left(\frac{\gamma c^2}{\lambda^2}\right)^2+...\right)\left( 1-\frac{V}{m\lambda^2}+\frac{1}{2}\left( \frac{V}{m\lambda^2}\right)^2+....\right)  \nn \\
&=& \sum_{j=0}^{\infty}  \frac{1}{j!} \left(\frac{-1}{m\lambda^2} \right)^{j-1} (\gamma mc^2 + V)^j \nn\\
&=& \sum_{j=0}^{\infty}  \frac{1}{j!} \left(\frac{-1}{m\lambda^2} \right)^{j-1} \mathcal{H}_{j,c}\;.
\end{eqnarray}
Finally, we will show that Hamiltonians in the infinite hierarchy also yield the same equation of equation. Using \eqref{42}, we find that
\begin{eqnarray}  -\frac{\partial \mathcal{H}_{j,c}}{\partial x} &=& \frac{d}{dt} \left(  \gamma m  \frac{\partial \mathcal{H}_{j,c}}{\partial p}  \right) \label{HJC}\;\nn\\
 - H_c^{j-1} \frac{dV}{dx}  &=&  H_{c}^{j-1} \dot p + p (j-1)\mathcal{H}_{c}^{j-2}\left(\frac{p}{\gamma m}\dot p + \dot x \frac{dV}{d x} \right)  \nn \\
 -\left( H_{c}^{j-1} + (j-1)H_{c}^{j-2}\frac{p^2}{\gamma m} \right) \frac{dV}{dx} &=& \left( H_{c}^{j-1} + (j-1)H_{c}^{j-2}\frac{p^2}{\gamma m} \right) \dot p     \nn \\
-\frac{d V}{d x} &=& \dot p \;, \nn
\end{eqnarray}
which is nothing but the desired equation of motion.

\section{Summary}
For the system with one degree of freedom, we successfully derive an alternative class, called the multiplicative form, of the Lagrangian $L_\lambda$ which is an extended class of the Lagrangian through the variable $\lambda$ both the relativistic case and non-relativistic case. A remarkable result here is that the multiplicative form of the Lagrangian is actually a generating function for an infinite hierarchy of Lagrangians producing the same equation of motion. To the best of our knowledge, this is the first time that a nontrivial set of Lagrangians, in terms of polynomial of kinetic and potential energies, has been systematically produced. Then the result in this paper again confirms a feature called the non-uniqueness of the Lagrangian apart from multiplying by a constant or adding the total time derivative term. 

The question is still open for the case of higher degrees of freedom. We do try to extend the idea to the case of two degrees of freedom by considering the system with the two identical particles interacting through an even potential function and succeed to construct the multiplicative form of the Lagrangian in appendix A. Another question is whether there are other classes of the Lagrangian, apart from these additive form and multiplicative form or not. If so, it may exist an universal form of the Lagrangian. This line of research is also worth pursuing.
One more remark that we would like to make is about the quantisation of the system. It is known that we can quantise the system through the Feynman path integrals for the standard additive form of the Lagrangian. One may ask what is the Feynman's quantisation method for the multiplicative Lagrangian. We will seriously answer this question elsewhere.
\appendix

\section{One dimensional two-particle system}\label{MB}
\numberwithin{equation}{section}
In this section, the idea is extended to the case of a system of two identical  particles in one dimension. The Hamiltonian of the system is given by
\begin{equation}\label{CM}
H_N(p_1,p_2,x_1,x_2)= \frac{{p_1}^2}{2m} + \frac{{p_2}^2}{2m} + V(x_1-x_2)\;,
\end{equation}
where $V(x_1-x_2)$ is the even function and may take the form of
\begin{eqnarray}
\delimiterfactor=1200 
V(x_1-x_2) = \left\{%
\begin{array}{ll}
g^2(x_1-x_2)^2 &\textrm{:Harmonic interaction },\\
\frac{g^2}{{(x_1-x_2)}^2} &\textrm{:Calogero-Moser intereaction \cite{CM}}.
\end{array}%
\right.
\end{eqnarray}
where $g$ is a coupling constant. The  equations  of motion for each particle reads
\begin{subequations}\label{EM}
\begin{eqnarray}
\ddot x_1&=&\frac{1}{m}\frac{\partial V(x_1-x_2)}{\partial x_1}\;,
\\  \ddot x_2 &=&-\frac{1}{m}\frac{\partial V(x_1-x_2)}{\partial x_2}
\;.
\end{eqnarray}
\end{subequations}
The additive Lagrangian of the system is
\begin{equation}
L_N(\dot{x_1},\dot{x_2},x_1,x_2) = \frac{1}{2}m\dot{x_1}^2  + \frac{1}{2}m\dot{x_2}^2 - V(x_1-x_2) \;, \label{AA5} 
\end{equation}
which can be obtained through Legendre transformation $L_N= p_1 \dot x_1 + p_2 \dot x_ 2-H_N$. To decouple the variables $x_1$ and $x_2$ in (5.3a) and (5.3b), the new set of variables, namely, $X=x_1+x_2$ and $x=x_1-x_2$ are used; and hence, (5.3a) and (5.3b) can be re-written as
\begin{subequations}\label{XX}
\begin{eqnarray}
\ddot X&=&0\;,\label{X1}\\
\ddot x&=&  - \frac{2}{m} \frac{dV(x)}{d x}\;.\label{X2}
\end{eqnarray}
\end{subequations}
Equation \eqref{X1} describes the motion of the centre of mass whereas \eqref{X2} describes the motion of the system in terms of the relative position between two particles.
\\\\
We are now looking for the multiplicative Lagrangian in terms of the variables $X$ and $x$ corresponding to (5.5a) and (5.5b).
Employing the result in the case of one particle, the multiplicative Lagrangian for two particles which describes the motion of free particle and the particle in the potential is
\begin{eqnarray}\label{LL2}
L_\lambda(\dot{X},\dot x,x) 
&=&  \frac{m\lambda^2}{2}\left( f(\dot X) + f(\dot x)g(x) \right)\; ,
\end{eqnarray}
where $f$ and $g$ are already defined in Section \ref{ML}. However, we repeatedly give them here again
  $$\begin{array}{rcl}
   f(\dot u) &=&  e^{-\frac{\dot{u}^2}{2\lambda^2}} + \frac{\dot{u}}{\lambda^2} \int_0^{\dot u} e^{-\frac{\xi^2}{2\lambda^2}} d \xi \; ,
\\ g(u) &=& e^{-\frac{2V(u)}{m\lambda^2}} \; . 
\end{array} $$
To see how the Lagrangian given by \eqref{LL2} leads to the equations of motion given by \eqref{XX}, we first put the Lagrangian into the Euler-Lagrange equation for the variable $X$
$$\begin{array}{rcl} \frac{\partial L_\lambda }{\partial X} - \frac{d}{dt}\left( \frac{\partial L_\lambda }{\partial \dot X} \right) &=& 0
\\ 0 - \frac{d}{dt} \left( \frac{d f(\dot X)}{d \dot X}\right) &=& 0
\\ \ddot X \left( m  \int_0^{\dot X} e^{\frac{-\xi^2}{2\lambda^2}} d\xi  \right) &=& 0\;.
\end{array}$$
This results in $\ddot X=0$. Next, we substitute the Lagrangian into the Euler-Lagrange equation for the variable $x$
$$\begin{array}{rcl}
 \frac{\partial L_\lambda }{\partial x} - \frac{d}{dt}\left( \frac{\partial L_\lambda }{\partial \dot x} \right) &=& 0
\\ f(\dot x)\frac{dg(x)}{d x} - \ddot{x} g(x) \frac{d^2 f(\dot x)}{{d \dot x }^2} -\dot x \frac{d g(x)}{dx} \frac{d f(\dot x)}{d \dot x} &=& 0 \\
f(\dot x) e^{-\frac{2V(x)}{m\lambda^2}}\left( \frac{-2}{m\lambda^2} \frac{d V(x)}{d x}\right) - \ddot x e^{-\frac{2V(x)}{m\lambda^2}} \frac{d^2 f(\dot x)}{{d \dot x}^2} \\
- \dot x e^{-\frac{2V(x)}{m\lambda^2}} \left( \frac{-2}{m\lambda^2} \frac{d V(x)}{d x}\right)\frac{d f(\dot x)}{d \dot x} &=& 0 
\\ \left( \frac{-2}{m\lambda^2} \frac{d V(x)}{d x}\right) \left( f(\dot x) -\dot x \frac{d f(\dot x)}{d \dot x}\right)-\ddot x \frac{d^2 f(\dot x)}{{d \dot x}^2} &=& 0 
\\  \left( \frac{-2}{m\lambda^2} \frac{d V(x)}{d x} - \frac{\ddot x}{\lambda^2} \right)\left( e^{-\frac{{\dot x}^2}{2\lambda^2}} \right) &=& 0
\\
\Rightarrow \ddot x &=&  - \frac{2}{m} \frac{dV(x)}{d x} \;,
\end{array}$$
which is indeed the equation of motion for $x$ variable.
\\
\\
In addition, in the limit that $\lambda$ approaches to infinity, the multiplicative Lagrangian becomes
$$\begin{array}{rcl}
\lim_{\lambda \rightarrow \infty}L_\lambda
&=& \lim_{\lambda \rightarrow \infty} \left[ m\lambda^2  \left(  1 + \frac{\dot X^2}{4\lambda^2}   + \frac{\dot x^2}{4\lambda^2} -   \frac{V(x)}{m\lambda^2} +\dots \right)  \right] \\

&=& \lim_{\lambda \rightarrow \infty} \left[  m\lambda^2   + \frac{m(\dot x_1 + \dot x_2)^2}{4}   + \frac{m(\dot x_1 - \dot x_2)^2}{4} -   V(x_1-x_2) +\dots  \right]\\

&=& \lim_{\lambda \rightarrow \infty} \left[  m\lambda^2   + \frac{m\dot x_1^2}{2}   + \frac{m\dot x_2^2}{2} -   V(x_1-x_2) +\dots  \right] \\
\lim_{\lambda \rightarrow \infty}\{ L_\lambda -m\lambda^2\}& =&  \frac{m\dot x_1^2}{2}   + \frac{m\dot x_2^2}{2} -   V(x_1-x_2) =L_N
\end{array}$$
which yields the additive Lagrangian.
\\\\
The multiplicative form of the Hamiltonian is given by
\begin{equation}\label{HSS}
H_\lambda 
 =   -\frac{m\lambda^2}{2} \left( k(P) + k(p)b(x)\right)\;,
  \end{equation}
where 
\begin{eqnarray}\label{HSS}
k(p)&=&e^{-\frac{p^2}{2m^2\lambda^2}}\;,\\
b(x)&=&e^{-\frac{V}{m\lambda^2}}\;,
\end{eqnarray}
and $ p = p_1 - p_2 $ and $P = p_1 + p_2$. The $p_1=m\dot x_1$ and $p_2=m\dot x_2$ are momenta for the primary and secondary particles, respectively. It is easy to show that the Hamiltonian given by \eqref{HSS} gives the equations of motion, i.e., \eqref{XX}. Firstly, we consider the equation of motion for the centre of mass
$$\begin{array}{rcl} 
 - \frac{\partial H_\lambda}{\partial X} &=&  m\frac{d}{dt} \left(\frac{\partial H_\lambda}{\partial P} \right)
\\ 0 &=& -\frac{m^2\lambda^2}{2} \frac{d}{dt} \left( \frac{d k(P) }{d P} \right)
\\ 0 &=& \dot P k(P) \;,
\end{array}$$
which $P$ is a constant. Secondly, we consider the equation for $x$ variable
$$\begin{array}{rcl}
- \frac{\partial H_\lambda}{\partial x} &=&  m\frac{d}{dt} \left(\frac{\partial H_\lambda}{\partial p} \right)
\\ -k(p)b(x)\frac{d V(x)}{dx} &=& \frac{1}{2} \frac{d}{dt} \left( p k(p)b(x) \right)
\\  &=&\frac{1}{2} \left(  pk(p)\frac{d b(x)}{dt} + b(x)\frac{d ( pk(p)) }{dt} \right)
\\ &=& \frac{1}{2} \left[ -\frac{2p^2k(p)b(x)}{m^2\lambda^2}  \frac{dV(x)}{d x} + b(x)\dot p \left(   -\frac{p^2}{m^2\lambda^2} k(p) + k(p)\right) \right]
\\
 -\frac{d V(x)}{d x}&=& \frac{1}{2}\left[   -\frac{2p^2}{m^2\lambda^2}\frac{d V}{d x} + \dot p \left(   -\frac{p^2}{m^2\lambda^2}  + 1\right) \right] 
\\ -2\frac{d V(x)}{d x}\left( -\frac{p^2}{m^2\lambda^2} +1\right) &=&   \dot p \left(   -\frac{p^2}{m^2\lambda^2}  + 1\right)\;\;\;\Rightarrow\;\;\; \dot p = -2\frac{d V(x)}{d x} \;,
\end{array}$$
which is again the equation of motion that we expected.
\\\\
Finally, we are interested to see how the multiplicative Hamiltonian behaves under the limit for very large $\lambda$ 
$$\begin{array}{rcl} 
\lim_{\lambda \rightarrow \infty } H_\lambda
 &=&  -\frac{m\lambda^2}{2}  \lim_{\lambda \rightarrow \infty }\left( 2 - \frac{P^2+p^2}{2m^2\lambda^2}  - \frac{2V}{m\lambda^2} + ... \right)\\
\lim_{\lambda \rightarrow \infty }\left( H_\lambda +m\lambda^2\right) &=& \frac{P^2+ p^2}{4m} + V = \frac{p_1^2}{2m} +\frac{p_2^2}{2m} + V  =H_N\;,
\end{array}$$
which is nothing but the standard Hamiltonian in the additive form. 
\\\\
 For the case of higher number of particles, especially the Calogero-Moser type systems \cite{Rin1,Rin2,Rin3}, there exists a hierarchy of the Lagrangians which are all in the additive form. Then it is interesting to see whether we could find the corresponding hierarchy of the multiplicative Lagrangians. 

\begin{acknowledgements}
Sikarin Yoo-Kong gratefully acknowledges the support from Theoretical and Computational Science Center (TaCS) under grant number: TaCS2558-2
\end{acknowledgements}

\end{document}